# Photoinduced Enhancement of the Charge Density Wave Amplitude

A. Singer[1,*], S. K. K. Patel[1,2], R. Kukreja[1,2], V. Uhlíř[2], J. Wingert[1], S. Festersen[3], D. Zhu[4], J. M. Glownia[4], H. Lemke[4,**], S. Nelson[4], M. Kozina[4], K. Rossnagel[3], M. Bauer[3], B. M. Murphy[5,3], O. M. Magnussen[3,5], E. E. Fullerton[2], and O. G. Shpyrko[1]

[1]Department of Physics, University of California-San Diego, La Jolla, California 92093, USA

[2]Center for Memory and Recording Research, University of California-San Diego, La Jolla, California 92093, USA

[3]Institute for Experimental and Applied Physics, University of Kiel, 24098 Kiel, Germany

[4]LCLS, SLAC National Accelerator Laboratory, Menlo Park, California 94025, USA

[5]Ruprecht Haensel Laboratory, Kiel University, 24098 Kiel, Germany

*Correspondence to: ansinger@ucsd.edu

**Present address: Paul Scherrer Institut, CH-5232 Villigen, Switzerland

Symmetry breaking and the emergence of order is one of the most fascinating phenomena in condensed matter physics. It leads to a plethora of intriguing ground states found in antiferromagnets, Mott insulators, superconductors, and density-wave systems. Exploiting states of matter far from equilibrium can provide even more striking routes to symmetry-lowered, ordered states. Here, we demonstrate for the case of elemental chromium that moderate ultrafast photo-excitation can transiently enhance the charge-density-wave (CDW) amplitude by up to 30% above its equilibrium value, while strong excitations lead to an oscillating, large-amplitude CDW state that persists above the equilibrium transition temperature. Both effects result from dynamic electron-phonon interactions, providing an efficient mechanism to selectively transform a broad excitation of the electronic order into a well defined, long-lived coherent lattice vibration. This mechanism may be exploited to transiently enhance order parameters in other systems with coupled degrees of freedom.

The coupling between various degrees of freedom drives the formation of complex orders in strongly correlated electron systems [1,2]. For instance, charge-lattice coupling creates charge-density-wave (CDW) order [3,4], while more intricate spin-charge-orbital-lattice coupling leads to combined charge and orbital order in manganites [5,6], or charge-spin stripe order in cuprates [2]. Recent studies suggest that spin order can be generated in iron pnictides in response to excitation of a coherent phonon [7], charge localization can be photoinduced in charge order systems [8], the superconducting order parameter in cuprates can be enhanced via suppression of the competing charge order or the transient redistribution of superconducting coherence [9–11], and hidden electronic states can be dynamically accessed [12–15]. Here we demonstrate a dramatic transient enhancement of the CDW amplitude in elemental chromium (Cr) following photo-excitation. This is remarkable because external excitation typically creates disorder, reduces the order parameter, and raises the symmetry [16–19]. We attribute the enhancement of the CDW amplitude to the dynamic electron-phonon interaction and experimentally discern multiple timescales, thus revealing the underlying physics.

The system we studied is a crystalline Cr film, which is antiferromagnetic and exhibits an incommensurate spin-density-wave (SDW) below the Néel temperature $T_N$=290±5 K. It also forms an incommensurate CDW, appearing as the second harmonic of the fundamental SDW ordering [3] (see Figs. 1 (a) and 1 (b)). The amplitude of the CDW can be directly measured by x-ray diffraction as the intensity of the corresponding satellite peaks. X-rays are mostly sensitive to the core electrons and the quantity we observe in our experiment is the elastic component of the CDW (periodic lattice distortion). Static x-ray data are presented in Fig. 1(c) and reveals that the CDW has a wave vector normal to the film surface, is pinned by the film surfaces, and is quantized with 8.5 periods in the film, as expected from earlier studies [3,20,21]. Here, the satellite peaks form constructive and destructive x-ray interference with the Laue oscillations, which leads to an increase of the scattered intensity at [0,0, 2-2δ] or decrease at [0,0, 2+2δ] [22,23] (2δ being the momentum transfer of the CDW). We used short optical laser pulses to excite ultrafast dynamics in the Cr thin film and the time dependent CDW amplitude was monitored via ultrafast x-ray diffraction (insets of Fig. 1(c)). The experiment was conducted at the x-ray free-electron laser at the Linac Coherent Light Source (LCLS) facility [26,27] in the stroboscopic mode. The sample thickness was 28 nm, the optical (x-ray) pulses had a wavelength of 800 nm (0.14 nm) and a pulse duration of 40 fs (15 fs), and the initial sample temperature was 115 K [23].

The time dependent x-ray diffraction signal of the CDW satellite peak at q=2-2δ is shown in Fig. 2 and reveals oscillations following photo-excitation (see [23] for Supplemental movies). The remarkable quality of the data allows unambiguous detection of four different time scales. The main oscillation has a period of 453±1 fs and is damped with a time constant of 3.0±0.5 ps so that about 20 oscillations are observed. Strikingly, for low pump fluences the mean of the oscillation increases rapidly in less than 0.5 ps, whereas for higher fluences it remains fixed at 0 (Fig. 2(b)). The CDW diffraction signal after significant damping of the oscillation (10 ps) decreases with pump fluence and reaches a value of 0 at 11 mJ/cm$^2$. At q=2+2δ an identically opposite behaviour occurs due to destructive interference [23] (see Fig. 2(b)). A positional shift of the Laue oscillations is observed for high fluences and shows a period of 8 ps (see Fig. 2(a)). This shift is initiated by the temperature increase of the lattice, and is used to calibrate the film temperature [23]. In the following we describe the physical processes that occur on shorter timescales.

The x-ray data measured at specific time delays within the first 0.5 ps are presented in Fig. 3 and the respective time dependent CDW is schematically depicted in the insets in Fig. 3(a). In the low-temperature equilibrium state (negative time delay), the x-ray data are consistent with the presence of the CDW nodes at both interfaces [22]. At a time delay of 0.11 ps after photo-excitation the dynamic x-ray data is in agreement with static x-ray data recorded above the Néel temperature $T_N$ and shows an undistorted lattice. At 0.22 ps we observed a reversal of the CDW amplitude, as revealed by the change of sign of the interference term in the diffraction signal [22,23]; the CDW still has nodes at the interfaces, however, the amplitude is inverted. The most striking observation of our study is the transient enhancement of the CDW amplitude at a time delay of 0.45 ps following moderate laser excitation (Figs. 2(b) and 3). The transient enhancement is oscillatory and occurs up to a time delay of 4 ps.

The fluence dependence of the pump-probe data reveals that the CDW amplitude is enhanced by about 30% above the maximum value in equilibrium at a fluence of 1 mJ/cm$^2$ (see Figs. 4(a)-(c)).

The experimental observation of the CDW amplitude enhancement was recorded in several independent measurements: the time traces for different fluences (Fig. 2), the fluence dependence of the dynamic CDW amplitude (Figs. 4, (a)-(c)), and in the measurements of both satellite peaks at $q=2-2\delta$ and $q=2+2\delta$ [23]. To exclude the possibility of exciting a broad phonon spectrum [28,29] we measured the transient signal at multiple q-values [23]. Only at the position of the satellite peak do we see an oscillation of the scattering intensity. This demonstrates that for low fluences no significantly excited lattice vibrations occur, apart from the CDW, and that the enhancement of the CDW amplitude is due to the pre-existing CDW in the film.

We attribute the enhancement of the CDW amplitude to dynamic electron-phonon interaction and present a model for the underlying physical processes in Fig. 4(d). The model is corroborated by fits to the data based on the theory of displacive excitation of coherent phonons [23,30–32]. We start with the low-temperature ground state (see Fig. 4(d), $\tau<0$ps), where the electronic order and electron-phonon coupling are responsible for the presence of the static CDW [3]. The schematic potential energy surface (blue curve) for the CDW amplitude, A, is centred at $A_0>0$, its value in the low-temperature ground state (see also Fig. 1(b)).

The photo-excitation creates hot charge carriers with temperatures well above $T_N$ within less than 50 fs [23,33], the electronic order is partially suppressed, and the electronic and lattice degrees of freedom are partially decoupled (see Fig. 4(d), $\tau\sim0$ps). The lattice distortion is still frozen, however, the potential energy surface has a new transient minimum at $A_1$ (red dashed) with a smaller or vanishing mean CDW amplitude due to the quenched electronic order. The lattice mode is thus released and starts oscillating. The frequency of this coherent lattice oscillation ($\nu=2.21$ THz) is in agreement with the frequency of the longitudinal acoustic phonon at the corresponding wavelength measured in bulk chromium [34]. The initial drop of the CDW amplitude after 0.22 ps saturates at a fluence of 4 mJ/cm$^2$ (Fig. 4 (c)) and A does not decrease below a value of -0.9. Therefore the amplitude $A_1=0$ in Fig. 4 (d) limits the initial displacement of the potential energy surface, indicating that only the pre-loaded energy due to the frozen phonon is released via quenching of the electronic order.

Surprisingly, for low fluences (smaller than 4 mJ/cm$^2$) we observe a dramatic deviation from the conventional model for displacive excitation of coherent phonons: here the minimum of the potential energy surface rapidly shifts back towards the initial ground state, $A_0$ (see Fig. 4(d), $\tau \sim t_{ep}$) [35]. To reproduce this essential feature of our data we added a superimposed exponential relaxation of the displacive component to the fit [23]. The time constant for this relaxation was determined to be 300 fs, in good agreement with the carrier-lattice thermalization time measured by optical reflectivity [33]. Therefore our analysis indicates that while the carriers cool down below $T_N$ the electronic ordering is re-established and pulls the quasi-equilibrium minimum of the potential energy surface towards higher values. In other words the electronic and the lattice degrees of freedom re-couple within less than 0.5 ps. This ultrafast backshift and the weak damping of the lattice oscillations about the dynamic quasi-equilibrium are the essence of the transient enhancement of the CDW amplitude (see Fig. 4(d), $\tau > t_{ep}$). The enhancement at 0.45 ps is maximized at a fluence of 1.2 mJ/cm$^2$ (see the vertical dashed line in Fig. 4(c)), which is coincident with the zero crossing of the CDW amplitude at 0.22 ps and indicates lattice-assisted re-condensation of the electronic ordering.

The relaxation time slightly increases with pump fluence, which is qualitatively supported by calculations within the two-temperature model [23,33,36–38], i.e. for increased fluences the carrier temperature stays longer above $T_N$. At even higher fluences (larger than 11 mJ/cm$^2$), when the quasi-equilibrium temperature after carrier-lattice equilibration approaches or exceeds $T_N$, no shift of the potential energy surface occurs and the CDW amplitude oscillates around zero, the value it would assume in equilibrium above $T_N$ (Fig. 2(b)). Since damping is also remarkably weak in the strong excitation regime, the CDW amplitude oscillations can persist above $T_N$. It is worth noting that the data measured with 1mJ/cm$^2$ in Fig. 2(b) could be interpreted in terms of an impulsive excitation of the coherent phonon mode, i.e. excitation without displacement of the potential energy surface, whose characteristic signature is a sine-type oscillatory behaviour [39]. The high fluence data (larger than 11 mJ/cm$^2$), however, clearly show that the coherent phonon is driven by a displacive excitation (cosine-type behaviour). The correct interpretation relies on the dynamical picture introduced in Fig. 4(d) and the re-forming of the electronic ordering is indispensable in explaining the CDW amplitude enhancement because the amplitude of the lattice oscillation is always smaller than the initial suppression of the mean value (see Table S1).

The damping time constant of the coherent lattice oscillation is about 3 ps and independent of the fluence. This surprisingly long time scale indicates anharmonic phonon-phonon interaction as the dominant decay channel. Electron-hole pair excitation, which typically leads to strong damping of order-parameter oscillations in strongly correlated electron systems [32,40,41], is expected to be ineffective here because of the SDW gap in the electronic spectrum [42]. Because of ultrafast carrier cooling and recondensation in the lattice distortion potential, this gap will quickly reopen, even after a complete quench of the electronic order [17]. Moreover, it is likely to persist above $T_N$ in the form of a pseudogap due to incipient magnetic order [42]. Finally, the period of the lattice oscillation is much shorter than the damping time and also does not depend on fluence [23]. Thus, the Cr system studied here essentially represents an effective converter of an electronic excitation into a well-defined, long-lived CDW amplitude oscillation: an oscillation that leads to a significant transient enhancement of the CDW amplitude and that can even persist above the equilibrium transition temperature. We anticipate that other sorts of excitation would lead to a similarly well-defined and persistent oscillation of the CDW amplitude in this system.

In summary, by using the unique capabilities now possible with the hard x-ray free-electron lasers we directly observe a dramatic enhancement of the CDW amplitude in chromium following photo-excitation: 30% above its maximum value in equilibrium. We identify the ultrafast underlying physical processes by discerning multiple timescales and explain our results by three main processes, referred to as "dynamic electron-phonon interaction" throughout the letter (I) the photo-induced quench of the electronic order unfreezes a coherent lattice oscillation (II) the mean amplitude of this lattice oscillation is increased due to the ultrafast re-condensation of the electronic order (III) the re-ordering of electrons is assisted by the still present lattice distortion. The rapid electronic re-condensation is evident from both the ultrafast backshift of the mean of the oscillation and the weak damping of the oscillation due to the reopening of the electronic gap. A further interesting question is whether the dynamic electron-phonon interaction can be combined with repeated photo-excitation to maintain the coherent lattice oscillation or to achieve an even higher enhancement of the CDW amplitude [43,44]. Our results also raise fundamental questions regarding the dynamics of the magnetic ordering and the electronic structure of the system. Finally, we anticipate that the enhancement of an order parameter via dynamic interaction of various degrees of freedom is a general phenomenon and can be observed and studied both theoretically and experimentally in a variety of systems including topological insulators [45] and strongly correlated electron materials [8].

**References:**

[1] M. Imada, A. Fujimori, and Y. Tokura, Rev. Mod. Phys. **70**, 1039 (1998).
[2] E. Dagotto, Science **309**, 257 (2005).
[3] E. Fawcett, Rev. Mod. Phys. **60**, 209 (1988).
[4] G. Gruner, *Density Waves In Solids* (Perseus Publishing, Cambridge, 2009).
[5] J. van den Brink, G. Khaliullin, and D. Khomskii, Phys. Rev. Lett. **83**, 5118 (1999).
[6] E. Dagotto, New J. Phys. **7**, 67 (2005).
[7] K. W. Kim, A. Pashkin, H. Schäfer, M. Beyer, M. Porer, T. Wolf, C. Bernhard, J. Demsar, R. Huber, and A. Leitenstorfer, Nat. Mater. **11**, 497 (2012).
[8] T. Ishikawa, Y. Sagae, Y. Naitoh, Y. Kawakami, H. Itoh, K. Yamamoto, K. Yakushi, H. Kishida, T. Sasaki, S. Ishihara, Y. Tanaka, K. Yonemitsu, and S. Iwai, Nat. Commun. **5**, 5528 (2014).
[9] D. Fausti, R. I. Tobey, N. Dean, S. Kaiser, A. Dienst, M. C. Hoffmann, S. Pyon, T. Takayama, H. Takagi, and A. Cavalleri, Science **331**, 189 (2011).
[10] R. Mankowsky, A. Subedi, M. Först, S. O. Mariager, M. Chollet, H. T. Lemke, J. S. Robinson, J. M. Glownia, M. P. Minitti, A. Frano, M. Fechner, N. A. Spaldin, T. Loew, B. Keimer, A. Georges, and A. Cavalleri, Nature **516**, 71 (2014).
[11] W. Hu, S. Kaiser, D. Nicoletti, C. R. Hunt, I. Gierz, M. C. Hoffmann, M. Le Tacon, T. Loew, B. Keimer, and A. Cavalleri, Nat. Mater. **13**, 705 (2014).
[12] K. Onda, S. Ogihara, K. Yonemitsu, N. Maeshima, T. Ishikawa, Y. Okimoto, X. Shao, Y. Nakano, H. Yamochi, G. Saito, and S.-Y. Koshihara, Phys. Rev. Lett. **101**, 067403 (2008).
[13] H. Ichikawa, S. Nozawa, T. Sato, A. Tomita, K. Ichiyanagi, M. Chollet, L. Guerin, N. Dean, A. Cavalleri, S. Adachi, T. Arima, H. Sawa, Y. Ogimoto, M. Nakamura, R. Tamaki, K. Miyano, and S. Koshihara, Nat. Mater. **10**, 101 (2011).
[14] L. Stojchevska, I. Vaskivskyi, T. Mertelj, P. Kusar, D. Svetin, S. Brazovskii, and D. Mihailovic, Science **344**, 177 (2014).
[15] V. R. Morrison, R. P. Chatelain, K. L. Tiwari, A. Hendaoui, A. Bruhács, M. Chaker, and B. J. Siwick, Science **346**, 445 (2014).
[16] F. Schmitt, P. S. Kirchmann, U. Bovensiepen, R. G. Moore, L. Rettig, M. Krenz, J.-H. Chu, N. Ru, L. Perfetti, D. H. Lu, M. Wolf, I. R. Fisher, and Z.-X. Shen, Science **321**, 1649 (2008).
[17] A. Tomeljak, H. Schäfer, D. Städter, M. Beyer, K. Biljakovic, and J. Demsar, Phys. Rev. Lett. **102**, 066404 (2009).
[18] M. Eichberger, H. Schäfer, M. Krumova, M. Beyer, J. Demsar, H. Berger, G. Moriena, G. Sciaini, and R. J. D. Miller, Nature **468**, 799 (2010).
[19] P. Beaud, A. Caviezel, S. O. Mariager, L. Rettig, G. Ingold, C. Dornes, S.-W. Huang, J. A. Johnson, M. Radovic, T. Huber, T. Kubacka, A. Ferrer, H. T. Lemke, M. Chollet, D. Zhu, J. M. Glownia, M. Sikorski, A. Robert, H. Wadati, M. Nakamura, M. Kawasaki, Y. Tokura, S. L. Johnson, and U. Staub, Nat. Mater. **13**, 923 (2014).
[20] H. Zabel, J. Phys. Condens. Matter **11**, 9303 (1999).
[21] E. E. Fullerton, J. L. Robertson, a. R. E. Prinsloo, H. L. Alberts, and S. D. Bader, Phys. Rev. Lett. **91**, 237201 (2003).
[22] A. Singer, M. J. Marsh, S. H. Dietze, V. Uhlíř, Y. Li, D. A. Walko, E. M. Dufresne, G. Srajer, M. P. Cosgriff, P. G. Evans, E. E. Fullerton, and O. G. Shpyrko, Phys. Rev. B **91**, 115134 (2015).
[23] See Supplemental Material at below for supplementary movies, further experimental

details, calculation of the CDW amplitude from the satellite peak intensity, data at q=2+2δ, details on x-ray data analysis, calculation of the fluence, calculation of the temperature, measurements at multiple q-values, Fourier spectra of the time traces, and fit details and results, which include Refs. [24,25].

[24]  S. Friedberg, I. Estermann, and J. Goldman, Phys. Rev. **85**, 375 (1952).
[25]  C. Kittel, *Introduction to Solid State Physics* (John Wiley & Sons, Inc, New York, 1953).
[26]  P. Emma, R. Akre, J. Arthur, R. Bionta, C. Bostedt, J. Bozek, A. Brachmann, P. Bucksbaum, R. Coffee, F.-J. Decker, Y. Ding, D. Dowell, S. Edstrom, A. Fisher, J. Frisch, S. Gilevich, J. Hastings, G. Hays, HeringPh., Z. Huang, R. Iverson, H. Loos, M. Messerschmidt, A. Miahnahri, S. Moeller, H.-D. Nuhn, G. Pile, D. Ratner, J. Rzepiela, D. Schultz, T. Smith, P. Stefan, H. Tompkins, J. Turner, J. Welch, W. White, J. Wu, G. Yocky, and J. Galayda, Nat Phot. **4**, 641 (2010).
[27]  D. Zhu, Y. Feng, S. Stoupin, S. A. Terentyev, H. T. Lemke, D. M. Fritz, M. Chollet, J. M. Glownia, R. Alonso-Mori, M. Sikorski, S. Song, T. B. van Driel, G. J. Williams, M. Messerschmidt, S. Boutet, V. D. Blank, Y. V Shvyd'ko, and A. Robert, Rev. Sci. Instrum. **85**, 063106 (2014).
[28]  M. Trigo, M. Fuchs, J. Chen, M. P. Jiang, M. Cammarata, S. Fahy, D. M. Fritz, K. Gaffney, S. Ghimire, A. Higginbotham, S. L. Johnson, M. E. Kozina, J. Larsson, H. Lemke, A. M. Lindenberg, G. Ndabashimiye, F. Quirin, K. Sokolowski-Tinten, C. Uher, G. Wang, J. S. Wark, D. Zhu, and D. A. Reis, Nat. Phys. **9**, 790 (2013).
[29]  D. Zhu, A. Robert, T. Henighan, H. T. Lemke, M. Chollet, J. M. Glownia, D. A. Reis, and M. Trigo, Phys. Rev. B **92**, 054303 (2015).
[30]  H. J. Zeiger, J. Vidal, T. K. Cheng, E. P. Ippen, G. Dresselhaus, and M. S. Dresselhaus, Phys. Rev. B **45**, 768 (1992).
[31]  A. Melnikov, I. Radu, A. Povolotskiy, T. Wehling, A. Lichtenstein, and U. Bovensiepen, J. Phys. D. Appl. Phys. **41**, 164004 (2008).
[32]  T. Huber, S. O. Mariager, A. Ferrer, H. Schäfer, J. A. Johnson, S. Grübel, A. Lübcke, L. Huber, T. Kubacka, C. Dornes, C. Laulhe, S. Ravy, G. Ingold, P. Beaud, J. Demsar, and S. L. Johnson, Phys. Rev. Lett. **113**, 026401 (2014).
[33]  S. D. Brorson, A. Kazeroonian, J. S. Moodera, D. W. Face, T. K. Cheng, E. P. Ippen, M. S. Dresselhaus, and G. Dresselhaus, Phys. Rev. Lett. **64**, 2172 (1990).
[34]  W. Shaw and L. Muhlestein, Phys. Rev. B **4**, 969 (1971).
[35]  The potential minimum does not reach its initial value A0 owing to the increase of the film temperature.
[36]  S. I. Anisimov, B. L. Kapeliovich, and T. L. Perelman, J. Exp. Theor. Phys. **66**, 375 (1974).
[37]  R. H. M. Groeneveld, R. Sprik, and A. Lagendijk, Phys. Rev. B **51**, 11433 (1995).
[38]  T. same qualitative result is obtained in the more sophisticated nonthermal electron model where the electron relaxation time is fluence independent and for increased fluences the carrier temperature remains longer above T. because of the higher initial Rise.
[39]  K. Ishioka and O. K. Misochko, *Progress in Ultrafast Intense Laser Science* (Springer Berlin Heidelberg, Berlin, Heidelberg, 2010).
[40]  P. Beaud, S. Johnson, A. Streun, R. Abela, D. Abramsohn, D. Grolimund, F. Krasniqi, T. Schmidt, V. Schlott, and G. Ingold, Phys. Rev. Lett. **99**, 174801 (2007).
[41]  H. Schaefer, V. V. Kabanov, and J. Demsar, Phys. Rev. B **89**, 045106 (2014).
[42]  R. Jaramillo, Y. Feng, J. C. Lang, Z. Islam, G. Srajer, H. M. Rønnow, P. B. Littlewood,

and T. F. Rosenbaum, Phys. Rev. B **77**, 184418 (2008).
[43] T. Onozaki, Y. Toda, S. Tanda, and R. Morita, Jpn. J. Appl. Phys. **46**, 870 (2007).
[44] P. Kusar, T. Mertelj, V. V. Kabanov, J.-H. Chu, I. R. Fisher, H. Berger, L. Forró, and D. Mihailovic, Phys. Rev. B **83**, 035104 (2011).
[45] I. Garate, Phys. Rev. Lett. **110**, 046402 (2013).

**Acknowledgements:**

The work at UCSD was supported by U.S. Department of Energy, Office of Science, Office of Basic Energy Sciences, under Contracts No. DE-SC0001805 (ultrafast x-ray scattering, A. S., R. K., J. W., and O. G. S.) and No. DE-SC0003678 (thin films synthesis and characterization, S. K. K. P., R. K., V. U., and E.E.F.). O.M.M., B.M.M., and S.F. thank the Bundesministerium für Bildung und Forschung (BMBF) for financial support via project 05K13FK2. Use of the Linac Coherent Light Source (LCLS), SLAC National Accelerator Laboratory, is supported by the U.S. Department of Energy, Office of Science, Office of Basic Energy Sciences under Contract No. DE-AC02-76SF00515.

# Figures

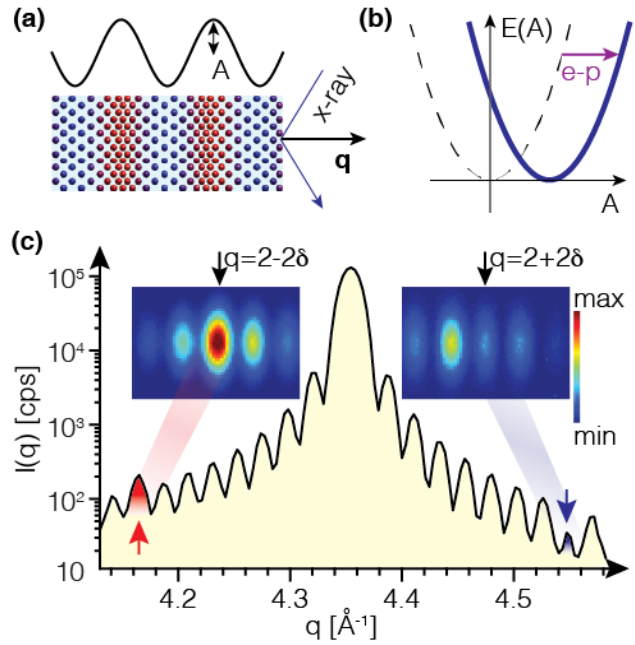

**Fig. 1: Static X-ray diffraction data. (a)** Schematic real space representation of the atomic positions in Cr in the presence of a CDW. The corresponding charge density modulation with the CDW amplitude, A, and the scattering geometry are also shown. **(b)** The potential energy surface for the CDW amplitude A. In the low temperature ground state the potential energy surface is shifted towards a non-vanishing value due to the electron phonon (e-p) coupling. **(c)** X-ray diffraction from a Cr thin film recorded with synchrotron radiation around the (002) Bragg peak (in photons per second) measured at a film temperature of 115 K. The intensity is increased (reduced) at the positions of the CDW satellites for low (high) q values (indicated by arrows). Insets: Diffraction patterns (linear scale) collected with the x-ray free-electron laser at two different momentum transfers $q=2-2\delta$ and $q=2+2\delta$ in the ground state, corresponding to different incident angles of x-rays.

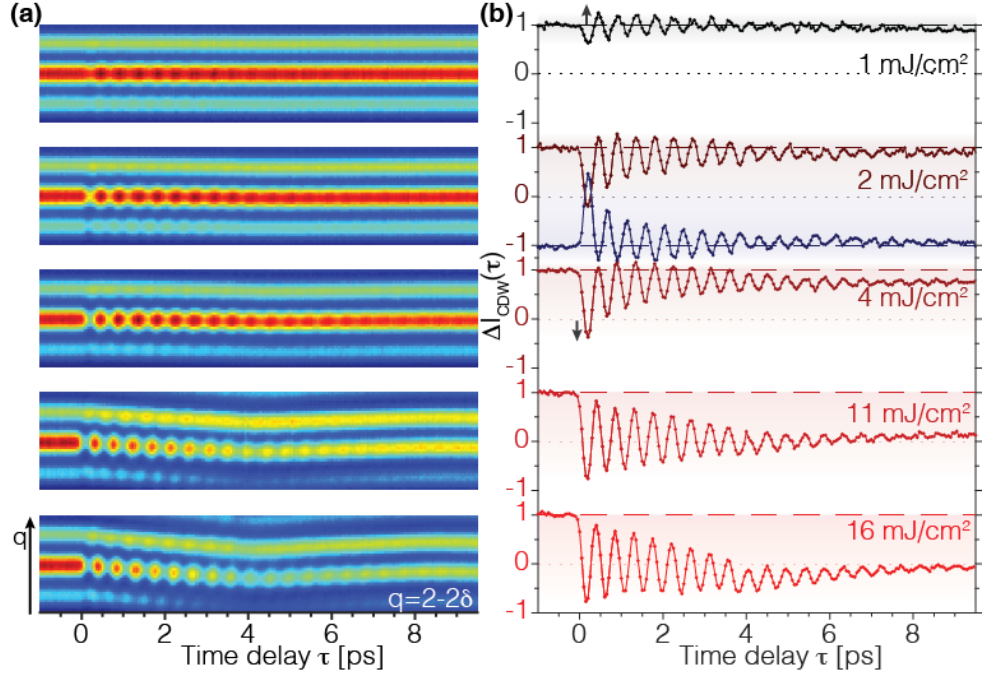

**Fig. 2: Time resolved x-ray diffraction data.** (a) Time dependence of the intensity in the vicinity of the CDW satellite at $q=2-2\delta$ (see Fig. 1) for a series of pump fluences (incident, p-polarization). (b) The black and red lines show the normalized transient intensity difference $\Delta I_{CDW}(\tau)=(I_{CDW}(\tau)-I_{RT})/|I_{CDW,0}-I_{RT}|$ at $q=2-2\delta$, where $I_{CDW}$ is the data in (a), $I_{RT}$ was measured at room temperature above $T_N$ without CDW, and $I_{CDW,0}$ was measured in the low temperature ground state. The intensity difference rises above its initial value of one for low fluences and drops below zero as the fluence increases (indicated by arrows). The blue line shows $\Delta I_{CDW}(\tau)$ at $q=2+2\delta$ for a fluence of 2 mJ/cm² and starts at -1 due to destructive interference.

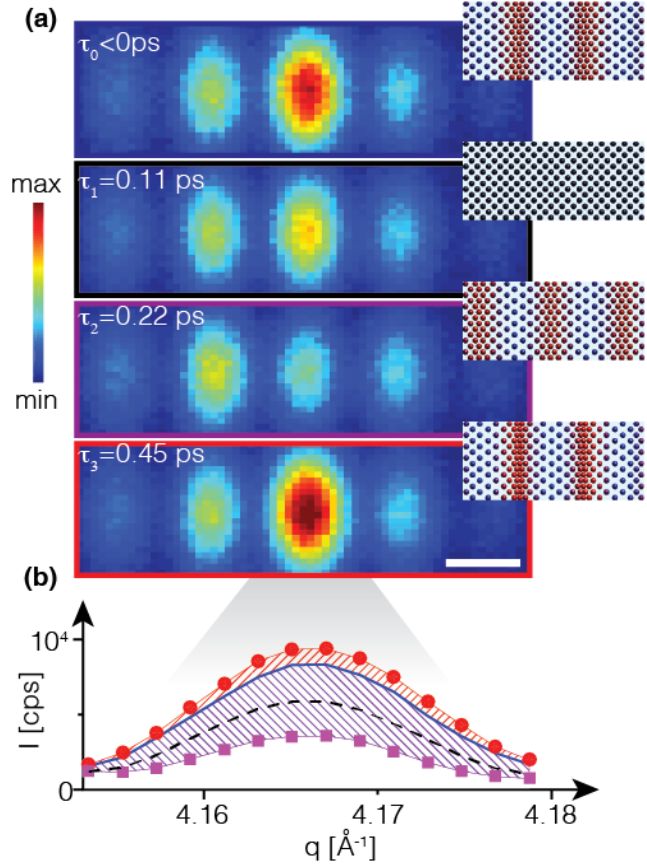

**Fig. 3: Interpretation of the time resolved x-ray diffraction data. (a,b)** X-ray data around q=2-2δ (**a**) and the integral along the vertical direction (**b**) in photons per second recorded at time delays $\tau_0$ before 0 ps (blue solid line), $\tau_1$=0.11 ps (black dashed line, 5 mJ/cm$^2$), $\tau_2$=0.22 ps (magenta squares, 5 mJ/cm$^2$), and $\tau_3$=0.45 ps (red circles, 1mJ/cm$^2$). The time delays represent significant instants in the first period of the oscillation in Fig. 2. **Insets in (a)** Schematic representations of the charge density modulation that are consistent with the x-ray data for different time delays. The charge density at 0.11 ps is similar to the room temperature case, where no CDW is present. The scale bar in **(a)** shows 0.025 Å$^{-1}$.

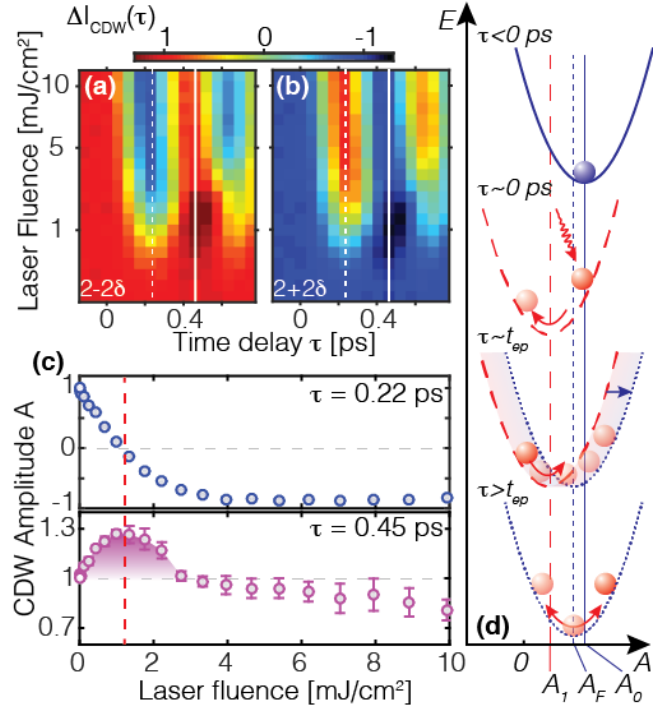

**Fig. 4: Enhancement of the CDW amplitude. (a,b)** The normalized intensity of the CDW satellite peak (see caption of Fig. 2) as a function of the pump fluence and time delay for $q=2-2\delta$ (**a**) and $q=2+2\delta$ (**b**). **(c)** The transient amplitude of the CDW at 0.22 ps (top, first minimum of the oscillation in Fig. 2) and 0.45 ps (bottom, first maximum of the oscillation in Fig. 2) extracted as line scans along the while lines in (**a**) and (**b**). **(d)** A schematic illustration of the mechanism behind the enhancement of the CDW amplitude due to dynamic electron-phonon interaction. The potential energy surfaces are drawn for the dynamic CDW amplitude, shown on the horizontal axis. $A_0$ is the amplitude of the CDW in the ground state, $A_1$ is the quasi equilibrium position after the excitation, and $A_F$ is its position after 10 ps. The transient amplitude $A(\tau)$ was fitted by the following equation $A(\tau)=A_F + B\cdot\cos(2\cdot\pi\cdot\tau'/\tau_P)\cdot\exp(-\tau'/\tau_D) - C\cdot\exp(-\tau'/\tau_{ep})$, where B is the amplitude, $\tau_P$ the period, $\tau_D$ the damping time of the oscillation, $\tau'=\tau-\tau_0$, $\tau_0$ is the offset of the oscillation, $C=A_F - A_1$, and $t_{ep}$ is the decay time for the shift of the quasi equilibrium towards $A_F$ [23]

# Photoinduced enhancement of the charge density wave amplitude.
# Supplementary Material.

A. Singer[1], S. K. K. Patel[1,2], R. Kukreja[1,2], V. Uhlíř[2], J. Wingert[1], S. Festersen[3], D. Zhu[4], J. M. Glownia[4], H. Lemke[4,**], S. Nelson[4], M. Kozina[4], K. Rossnagel[3], M. Bauer[3], B. M. Murphy[5,3], O. M. Magnussen[3,5], E. E. Fullerton[2], and O. G. Shpyrko[1]

[1]Department of Physics, University of California-San Diego, La Jolla, California 92093, USA

[2]Center for Memory and Recording Research, University of California-San Diego, La Jolla, California 92093, USA

[3]Institute for Experimental and Applied Physics, Kiel University, 24098 Kiel, Germany

[4]LCLS, SLAC National Accelerator Laboratory, Menlo Park, California 94025, USA

[5]Ruprecht Haensel Laboratory, Kiel University, 24098 Kiel, Germany

*Correspondence to: ansinger@ucsd.edu

**Present address: Paul Scherrer Institut, CH-5232 Villigen, Switzerland

**Animated time resolved x-ray data**

https://www.youtube.com/watch?v=3QMhuDv8ihs
https://www.youtube.com/watch?v=XkO8GAVmvFw
https://www.youtube.com/watch?v=GAVKxCr0cdM
https://www.youtube.com/watch?v=QFwnG0M_u68
https://www.youtube.com/watch?v=_ir08gcQmHs

All movies show the dynamic x-ray data as measured on the 2D detector (scattering vector q is horizontal) and the projection onto q. The time delay between the optical pump and x-ray probe is indicated in the top right corner.

**Experimental details**

The thin chromium film was deposited onto the single-crystal MgO(001) substrate using DC magnetron sputtering at a substrate temperature of 500° C and annealed for 1 h at 800° C. The growth process was optimized to yield both a smooth surface and good crystal quality of the sample. The film thickness was determined to be 28 nm by x-ray diffraction at the Advanced Photon Source (Fig. 1(c)). The pump-probe experiment was carried out at the XPP instrument of the LCLS with an x-ray photon energy of 8.9 keV, selected by the (111) diffraction of a diamond crystal. X-ray diffraction in the vicinity of the out of plane (002) Bragg peak (2θ=60 degrees) from each pulse was recorded by an area detector (CS140k) with a repetition rate of 120 Hz. Due to the mosaic spread of the crystal in the film plane, a number of Laue oscillations are observed on the area detector simultaneously. About 100 pulses were recorded for each time delay (50 fs steps in the time traces). For every time delay separately, the intensity was dark noise corrected and normalized by the intensity measured in the region of the area detector where Laue oscillations were absent. The sample was excited by optical (800 nm, 40-fs), p-polarized laser pulses propagating approximately collinearly with the x-ray pulses. The final temporal resolution was estimated at 80 fs. The spot sizes (full width at half maximum) of the optical and x-ray pulses were 0.46 mm (H) x 0.56 mm (V) and 0.2 mm (H) x 0.2 mm (V), respectively.

**Intensity at the CDW satellite position**
It can be readily shown [1] that in the presence of a CDW the positions of the atomic planes $r_n$ can be written as (normal to the CDW wave vector)
$$r_n = a \cdot n + A \cos(2\delta a n - \phi_0),$$
where $n$ is an integer, $A$ is the CDW amplitude, $2\delta$ is the scattering vector corresponding to the CDW periodicity, $a$ is the lattice parameter, and $\phi_0$ is the relative phase of the CDW modulation with respect to the substrate interface. In an x-ray experiment with a momentum transfer $q$ normal to the planes $r_n$ the scattered intensity can be written as [2]
$$I(q) = I_N(|f(q)|^2 + qA \sin(\alpha)[f(q)f(q-2\delta) - f(q)f(q+2\delta)] + \frac{(qA)^2}{4}[f(q-2\delta)^2 + f(q+2\delta)^2]), \quad (1)$$
where $q$ is the momentum transfer, $I_N$ is a normalization constant, $f(q) = \sin(qaN)/\sin(qa)$, $N$ is the number of atomic layers in the film, and $\alpha = N_P\pi - \phi_0$, with $N_P$ being the number of CDW periods in the film. In our system, $\phi_0 \sim \pi$ and $N_P = 8.5$. For bulk and relatively thick crystals only the first and the last terms are typically observed since $f(q)$ is extremely sharp. Here, $f(q)$ is relatively broad due to the small thickness of the film, the last term can be neglected, and the second term (interference term) dominates. Equation (1) shows that the measured intensity is enhanced or reduced in the presence of the CDW, identically opposite on both satellites $q \pm 2\delta$ around the Bragg peak, and that the amplitude $A$ of the CDW is directly proportional to the intensity of the interference terms, as discussed in the main text.

**Analysis of the time resolved x-ray data**
The signal recorded on the area detector is not identical to a typical θ-2θ scan shown in Fig. 1(c); instead it represents a portion of the Ewald sphere through the Bragg rod, which is oriented normal to the film surface (insets in Fig. 1(c)). A correction was required in order to be able to subtract the intensity measured at room temperature (without the CDW), which due to lattice expansion showed shifted Laue oscillations, and to correct for the positional shift of the Laue oscillations due to the coherent phonon at q=0 for high fluences. During this correction we normalized the diffraction data by a

Gaussian function, whose width was determined from the data in the direction perpendicular to $q_{002}$ (vertical direction in the inset of Fig. 1(c)) and rescaled according to experimental geometry. The center of this Gaussian function was found by comparing the corrected data with a true θ-2θ scan collected at a synchrotron. The finally determined CDW amplitude was not sensitive to the details of the procedure and a shift of the center of the Gaussian function over a few pixels had a negligible effect on the results. For final analysis data similar to the inset in Fig. 1(c) was projected vertically. Figures 2 and 4 present the behavior of the satellite peak, which was calculated as an average over 10 pixels around the peak center (see also Fig. 3(a)). The CDW amplitude shown in Fig. 2(b) and 4(a-c) was calculated using equation (1) from the measured intensity of the satellite peak after photo-excitation and the intensity of the respective Laue oscillation above the Néel temperature $T_N$ in absence of the CDW.

**Calculation of the fluence**
The fluence of the incident laser pulses was controlled by the angle of a wave plate. In particular for low angles of the wave plate we measured a non-vanishing transmitted intensity, indicating additional non p-polarized components, which have a different absorption coefficient. Additionally, possible drifts of the laser beam or the x-ray beam could compromise the spatial overlap and the sample at the tails of the laser beam could be probed, leading to in fact smaller excitation fluences. Importantly, due to collinear geometry such drifts do not compromise the temporal overlap. To avoid these systematic uncertainties we calibrated the fluence of the incident laser pulses with the transient shift of the Bragg peak, measured for each of the curves in Fig. 2 independently at 100 ps. This Bragg peak shift is in good agreement with the shift of the Laue oscillations, which is slightly smaller / larger at q=2-2δ / q=2+2δ due to film thickness increase upon lattice expansion. Assuming the linear expansion coefficient is constant above 115 K (measured, not shown here) and the heat capacity is constant (calculated using the Debye model, not shown here), the fluence was determined as $F = \Delta \cdot F_N/\Delta_N$, where $F$ is the fluence, $\Delta$ is the Bragg peak shift, and the subscript $N$ denotes the normalization values, which were measured with a wave plate angle of 10 degrees.

**Fitting of the transient CDW amplitude**
The transient amplitude $A(\tau)$ was fitted by the following equation [3]

$$A(\tau) = A_F + B \cos\left(2 \cdot \pi \cdot \frac{\tau - \tau_0}{t_P}\right) \cdot \exp\left(-\frac{\tau - \tau_0}{t_D}\right) - C \cdot \exp\left(-\frac{\tau - \tau_0}{t_{ep}}\right)$$

where $A_F$ is the final amplitude of the CDW at 10 ps, $B$ is the amplitude, $t_P$ the period, and $t_D$ the damping time of the oscillation. The parameter $\tau_0$ is the offset of the oscillation, $C = A_F - A_1$ (see Fig. 4(d)), and $t_{ep}$ is the decay time for the shift of the quasi equilibrium towards $A_F$. The last term is indispensable to accurately reproduce the data. The fit results are presented in Fig. S1 and Table S1.

**Calculation of the temperature**

We have simulated the electron $T_e$ and lattice $T_l$ temperatures by solving the coupled differential equations within the two temperature model [2,4]

$$C_e(T_e)\frac{dT_e}{dt} = -G(T_e - T_l) + S(t)$$

$$C_l(T_l)\frac{dT_l}{dt} = G(T_e - T_l),$$

where $G=4.6\times10^{11}$ W/(cm$^3$K) is the electron-phonon coupling constant [5,6], $C_e=\gamma T_e$ with $\gamma=211$ J/(m$^3$K) is the electron heat capacity [6], $C_l$ is the lattice heat capacity, which was calculated within the Debye approximation [7], $t$ denotes time, and $S(t)$ is the IR laser excitation, which was simulated as a Gaussian with 40 fs FWHM with a height adjusted to yield the final film temperature. This temperature was calculated by comparing the transient Bragg peak position after 100 ps with the thermal expansion coefficient (measured on the same sample). The inability of the film to expand laterally was accounted for by correcting the observed lattice expansion in the pump probe experiment by 1+2$\nu_P$, where $\nu_P=0.29$ is the Poisson ratio. At 250 K the CDW with the initial periodicity vanishes for temperatures lower than $T_N$, and a CDW with a different periodicity emerges. The sample is optically thin and homogeneously heated, thus the heat transport during the calculated time period can be neglected [5].

| Momentum transfer | Fluence [mJ/cm$^2$] | Period [ps] | $A_F$ [arb. u.] | $\tau_{ep}$ [ps] | $\tau_d$ [ps] | B/(1-$A_F$+C) |
|---|---|---|---|---|---|---|
| q=2-2δ | 1 | 0.453±0.002 | 0.99±0.01 | 0.14±0.05 | 2.85±0.5 | 0.64±0.10 |
| q=2-2δ | 2 | 0.454±0.001 | 0.87±0.01 | 0.38±0.04 | 3.00±0.2 | 0.68±0.05 |
| q=2-2δ | 4 | 0.453±0.001 | 0.76±0.01 | 0.41±0.03 | 2.95±0.2 | 0.66±0.03 |
| q=2-2δ | 11 | 0.453±0.001 | 0.03±0.01 | 0.02±0.43 | 3.07±0.3 | 0.86±1.65 |
| q=2+2δ | 1 | 0.454±0.002 | 0.93±0.01 | 0.23±0.07 | 2.95±0.4 | 0.73±0.20 |
| q=2+2δ | 2 | 0.454±0.001 | 0.95±0.01 | 0.34±0.05 | 2.27±0.2 | 0.69±0.11 |
| q=2+2δ | 4 | 0.452±0.001 | 0.82±0.01 | 0.48±0.05 | 2.62±0.2 | 0.73±0.08 |
| q=2+2δ | 11 | 0.452±0.001 | 0.01±0.01 | 0.05±0.02 | 2.68±0.1 | 0.67±0.13 |

**Table S1:** Parameters determined from the fit to the data for time delays between 0 and 5 ps. $A_F$ is the final mean amplitude of the CDW, $t_{ep}$ is the time scale of the shift of the oscillation (solid line in Fig. S1), $t_d$ is the damping time of the oscillation. The last column represents the ratio between the amplitude of the oscillation, **B**, and the initial suppression of the CDW amplitude **1-$A_F$+C**. The upper and lower part of the table represent the transient behavior observed at q=2-2δ and q=2+2δ respectively. The fitting procedure for the highest fluence (16 mJ/cm$^2$) was compromised by the coherent phonon at the zone boundary and the parameters are not shown.

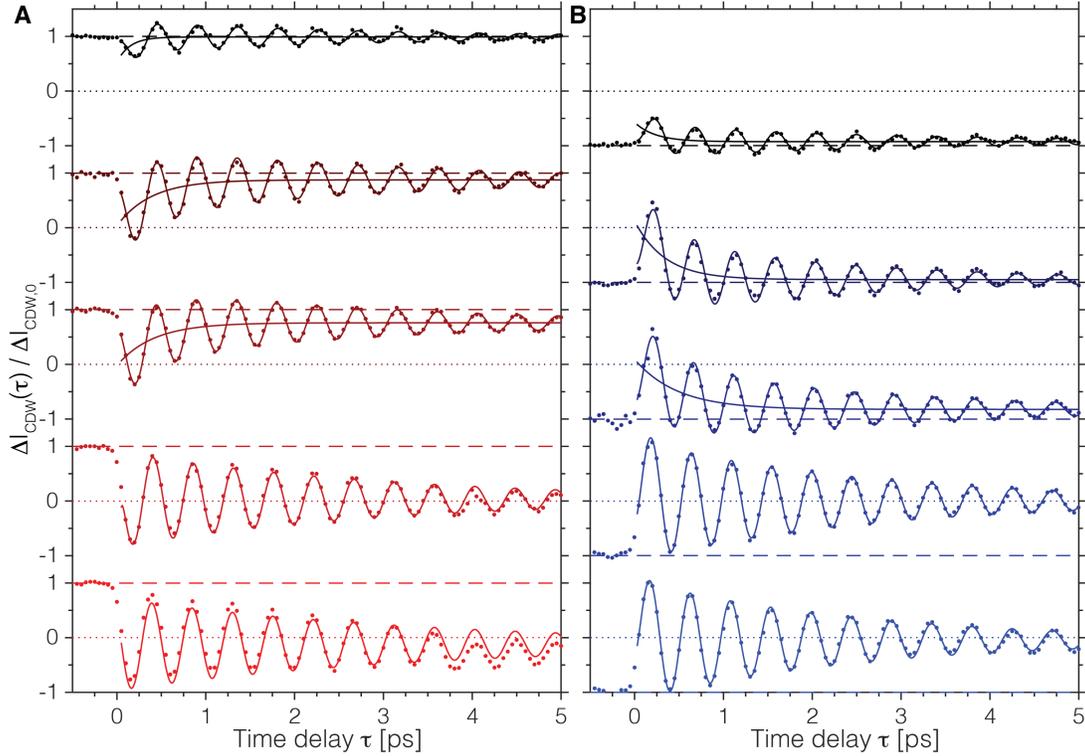

**Fig. S1:** Fits to the data using the model described in methods for q=2-2δ (**A**) and q=2+2δ (**B**). Fluences from top to bottom: 1 mJ/cm$^2$, 2 mJ/cm$^2$, 4 mJ/cm$^2$, 11 mJ/cm$^2$, 16 mJ/cm$^2$. The fitting procedure for the highest fluence (16 mJ/cm$^2$) was compromised by the coherent phonon at the zone boundary.

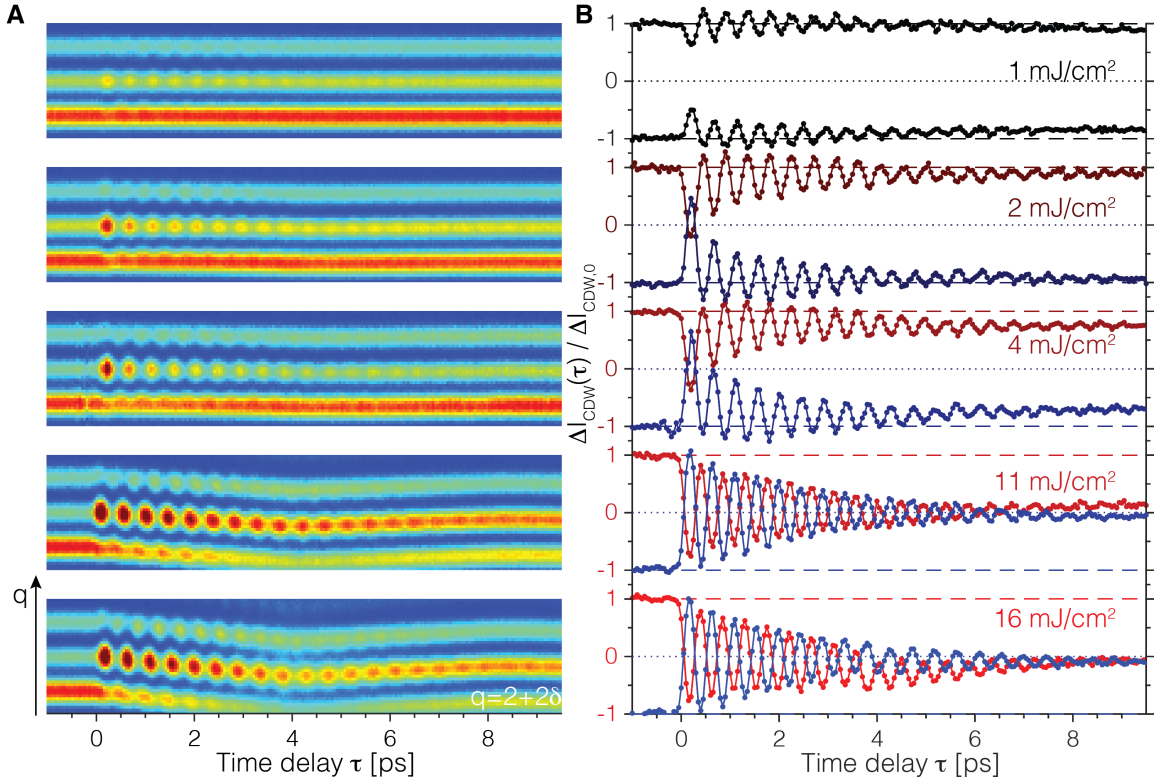

**Fig. S2:** (**A**) Same as Fig. 2 of the main text measured for q=2+2δ. (**B**) The normalized transient intensity difference $\Delta I_{CDW}(\tau)$ at q=2+2δ (black-blue) and q=2-2δ (black-red).

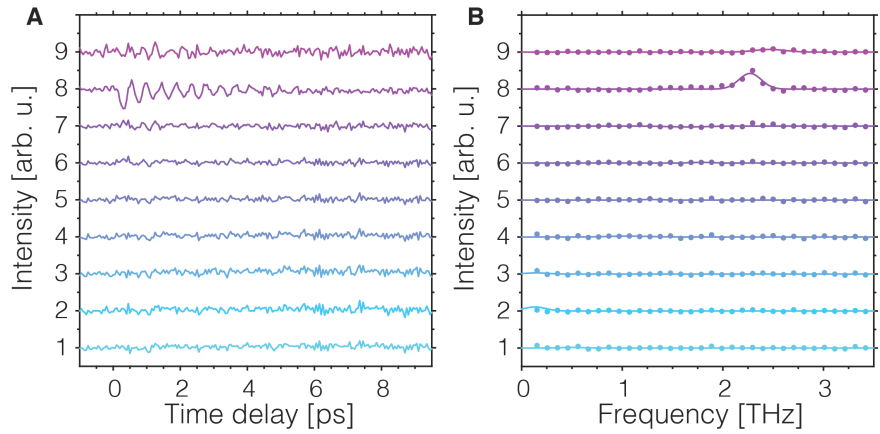

**Fig. S3:** (**A**) The time trace of the diffracted intensity measured with a pump fluence of 1 mJ/cm² at different q values shifted vertically for better visibility and labeled by the Laue oscillation counted from the Bragg peak (see Fig. 1 (**c**)). The CDW satellite is on fringe 8. (**B**) Fourier spectra of (**A**) taken in the range from 0.5 to 9.5 ps.

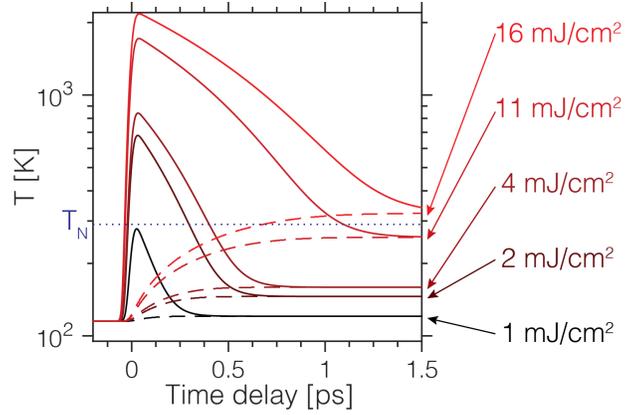

**Fig. S4:** Carrier (solid lines) and lattice (dashed lines) temperatures calculated within the two-temperature model with parameters identical to [2].

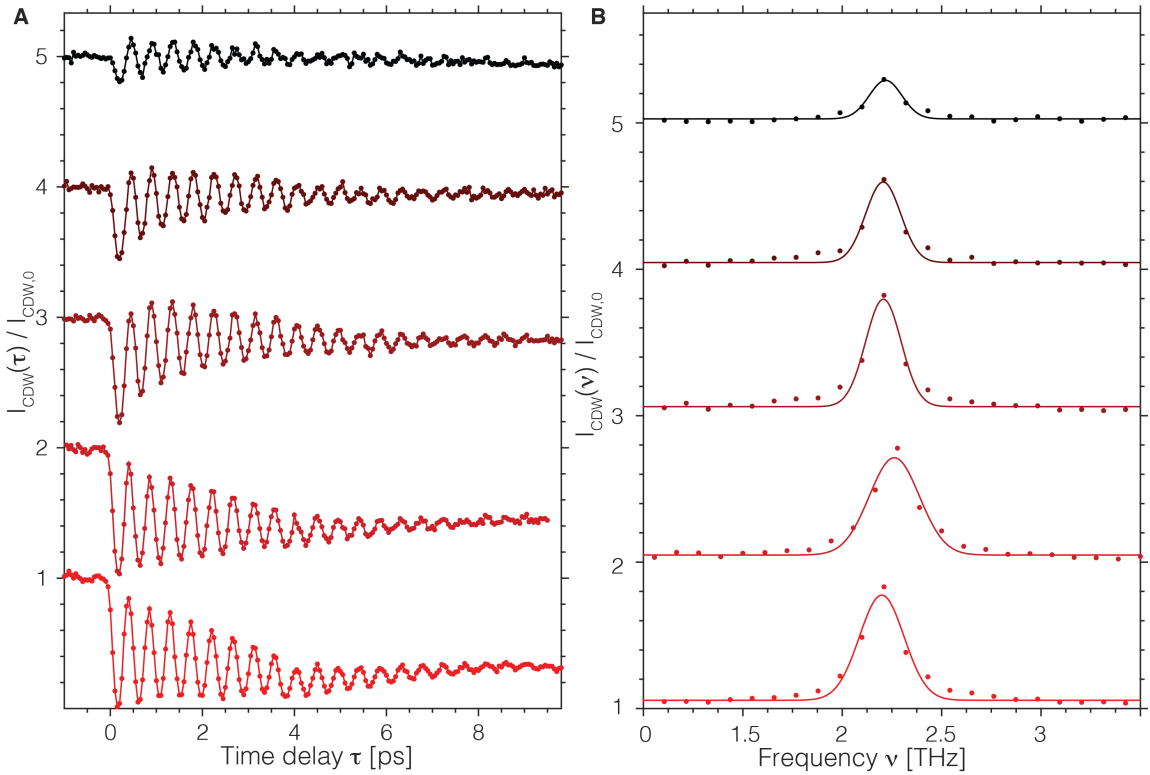

**Fig. S5:** **(A)** The time trace of the diffracted intensity measured at q=2-2δ. **(B)** Fourier spectra of **(A)** taken from 0.5 to 9.5 ps. Fluences from top to bottom as in Figure S1.

[1]  E. Fawcett, Rev. Mod. Phys. **60**, 209 (1988).
[2]  A. Singer, M. J. Marsh, S. H. Dietze, V. Uhlíř, Y. Li, D. A. Walko, E. M. Dufresne, G. Srajer, M. P. Cosgriff, P. G. Evans, E. E. Fullerton, and O. G. Shpyrko, Phys. Rev. B **91**, 115134 (2015).
[3]  H. J. Zeiger, J. Vidal, T. K. Cheng, E. P. Ippen, G. Dresselhaus, and M. S. Dresselhaus, Phys. Rev. B **45**, 768 (1992).
[4]  S. I. Anisimov, B. L. Kapeliovich, and T. L. Perelman, J. Exp. Theor. Phys. **66**, 375 (1974).
[5]  S. D. Brorson, A. Kazeroonian, J. S. Moodera, D. W. Face, T. K. Cheng, E. P. Ippen, M. S. Dresselhaus, and G. Dresselhaus, Phys. Rev. Lett. **64**, 2172 (1990).
[6]  S. Friedberg, I. Estermann, and J. Goldman, Phys. Rev. **85**, 375 (1952).
[7]  C. Kittel, *Introduction to Solid State Physics* (John Wiley & Sons, Inc, New York, 1953).